# MANAGEMENT LANGUAGE SPECIFICATIONS
# FOR DIGITAL ECOSYSTEMS

Youssef Bassil

LACSC – Lebanese Association for Computational Sciences
Registered under No. 957, 2011, Beirut, Lebanon
youssef.bassil@lacsc.org

*Abstract:* This paper defines the specifications of a management language intended to automate the control and administration of various service components connected to a digital ecosystem. It is called EML short for Ecosystem Management Language and it is based on proprietary syntax and notation and contains a set of managerial commands issued by the system's administrator via a command console. Additionally, EML is shipped with a collection of self-adaptation procedures called SAP. Their purpose is to provide self-adaptation properties to the ecosystem allowing it to self-optimize itself based on the state of its execution environment. On top of that, there exists the EMU short for Ecosystem Management Unit which interprets, validates, parses, and executes EML commands and SAP procedures. Future research can improve upon EML so much so that it can be extended to support a larger set of commands in addition to a larger set of SAP procedures.

## INTRODUCTION

Today, with the booming of information technologies and advances in computing field, information systems and IT infrastructures need to implement an agile set-up in which dynamic business and organization evolution are the key. It is with no doubt that the latest accomplishments in B2B (Business-to-Business) implementations are permitting large enterprises to accelerate the dynamic of business and tackle such challenges. In practice, service science is the major contributor in this new paradigm as it focuses on building component-based across-enterprise business models that can cope with the ever-changing business constraints, trends, and requirements [1]. Recently, a more forward-thinking and evolved architecture has been adopted by the computer and information society, it is called Digital Ecosystem or DE for short. A digital ecosystem is a distributed IT infrastructure built using interrelated e-service models [2] that exhibit such properties as sustainability, standardization, self-organization, self-management, self-integration, and self-adaptation [3], [4]. It is inspired by natural ecosystems that evolve and adapt according to their living environment. In effect, digital ecosystems allow building business models for sophisticated, distributed, and collaborative e-enterprises, e-marketplaces, e-communities, and e-cities using reusable service components [5].

This paper proposes a management language for digital ecosystems called EML short for Ecosystem Management Language. It is a proprietary language based on proprietary syntax and notation used to manage and control the different service components connected to the digital ecosystem. EML is powered by the EMU short for Ecosystem Management Unit which houses the EML interpreter that decodes and executes EML commands. The EML language offers a set of managerial commands whose scope includes but not limited to integrating services, updating and deleting service WSDLs, retrieving service details, granting and revoking security access rights, and monitoring and reporting. Furthermore, the EML language features a set of

Self-Adaptation Procedures (SAP) whose purpose is to add self-adaptation properties to the ecosystem. The scope of SAP includes but not limited to dynamically allocating memory space, dynamically reducing power consumption, dynamically allocating CPU cores, and dynamically allocating printer devices. All in all, EML is aimed at automating and easing the management and administration of various service modules interconnected within a digital ecosystem.

## RELATED WORK

Little work has been done to develop a standard management language for digital ecosystems. A sole attempt is the OASIS reference model [6] which is a generic framework for building and managing service-oriented architectures. It is majorly composed of six units: the orchestration and management unit which is responsible for administering the connected components and web services in the SOA; the data content unit which represents a set of databases that feed web services with data and information; the service description unit which defines the functions exposed by the connected web services in the SOA; the service discovery unit which contains a look-up registry to locate and consume web services; the messaging unit which can be thought as the communication medium that lets all connected components share data and communicate between each other; and the security and access unit which provides a security layer for securing and encrypting the messages being sent and received between the different components of the SOA. Figure 1 depicts the different building blocks of the OASIS reference model.

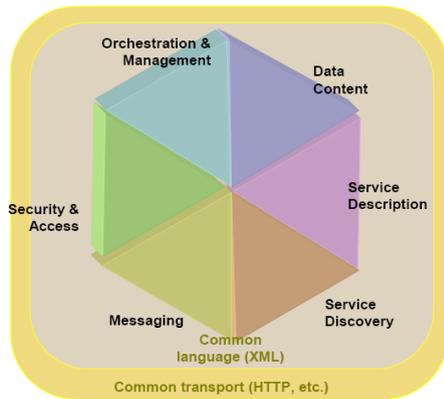

Figure 1. OASIS model

As for the OASIS management unit, it defines a combination of manual and automated management for service components that are controlled and proactively monitored with respect to the business context. It is a management process for controlling SOA ecosystem resources according to the policies and principles defined by the system's governance. These resources include services, service descriptions, service contracts, policies, roles, security, and people, in addition to the business relationships between them. Moreover, OASIS defines several domains of interest within the management framework, they include: the management and control of service resources that are connected to any complex structure; the declaration and enforcement of service contracts and policies approved by the stakeholders of the SOA ecosystem; and the management of the relationships between the different participants that use and offer services to each other. According to OASIS, services are managed using their metadata which are a set of properties and attributes pertaining to any produced or consumed service within the ecosystem. There exist four types of managements: Configuration management which controls the configuration of the deployment of new services into the ecosystem; event monitoring management which allows managing the execution of particular service functionalities; performance management which controls service results and their effects against the business goals and objectives of the service; management of quality of service (QoS) which manages the service non-functional characteristics associated with service quality; and policy management which allows adding, deleting, and modifying systems policies associated with the SOA ecosystem.

## EML – ECOSYSTEM MANAGEMENT LANGUAGE

The proposed Ecosystem Management Language (EML) is a declarative language based on proprietary syntax and notation used to control and to administer service components connected to the digital ecosystem infrastructure. The purpose of the EML is to ease and automate the management and administration of different entities in the ecosystem. Its scope includes integrating and disintegrating services, disabling and enabling existing services, updating and deleting services' WSDLs, monitoring and reporting, creating and deleting service replicas, and resource access control.

In essence, the EML language is sub-divided into several language elements including: The *command-name* denoting the type of command to be executed by the EML interpreter, one or more parameters (*param1, param2, paramN*) denoting data to be passed to the addressed service, and an *acknowledge-command* denoting whether or not a given command was successful. More formally, it can be represented as follows:

*command-name: param1, param2, paramN*
*acknowledge-command: param1, param2, paramM, True|False*

The core of the EML language is an EMU unit short for Ecosystem Management Unit that houses the EML interpreter which scans an issued EML command, extracts valuable tokens out if it, parses them to validate their correct arrangement, and then executes the command over the operational ecosystem. Figure 2 depicts the EMU along with the built-in EML interpreter.

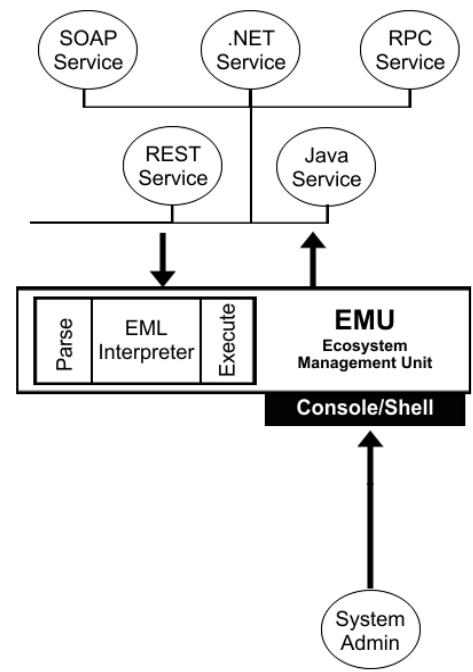

Figure 2. EMU and its EML interpreter

The EML is capable of several managerial operations via a set of commands including: *bind*, *unbind*, *update*, *delete*, *enable*, *getClients*, *grant*, *createReplica*, *getInfo*, and *executeSAP*. Their complete specifications are described below:

*bind: serviceURL, WSDL*
*bind-ack: service-ID, True|False*

The "bind" command is used to integrate (connect) a new service into the current ecosystem. The system's administrator initiates a "bind" command sending as parameters the URL and the WSDL of the new service. The EML environment replies back with a "bind-ack" command that contains the ID (auto-generated by the EML environment) of the service just integrated and a boolean value indicating whether or not the operation was successful. In case a false acknowledgment was returned, service-ID parameter would be equal to -1; whereas, in case a true acknowledgment was returned, the ID, IP, URL, and WSDL of the new service would be stored in a discovery registry database for later reference.

*unbind: service-ID*
*unbind-ack: service-ID, True|False*

The "unbind" command is used to disintegrate (disconnect) an existing service from the ecosystem. The system's administrator initiates a "unbind" command sending as parameter the ID of the service to be disintegrated. The EML environment replies back with an "unbind-ack" command that contains the ID of the service just disintegrated and a boolean value indicating whether or not the operation was successful.

*update: service-ID, WSDL*
*update-ack: service-ID, True|False*

The "update" command is used to update the WSDL of an existing service. The system's administrator initiates a "update" command sending as parameters the ID of the service whose WSDL is to be updated and the actual new WSDL. The EML environment replies back with an "update-ack" command that contains the ID of the addressed service and a boolean value indicating whether or not the WSDL was updated successfully.

*delete: service-ID*
*delete-ack: service-ID, True|False*

The "delete" command is used to delete the WSDL of an existing service. The system's administrator initiates a "delete" command sending as parameter the ID of the service whose WSDL is to be deleted. The EML environment replies back with a "delete-ack" command that contains the ID of the addressed service and a boolean value indicating whether or not the WSDL was deleted successfully.

*enable: service-ID, True|False*
*enable-ack: service-ID, True|False*

The "enable" command is used to enable or disable an existing service. The system's administrator initiates an "enable" command sending as parameters the ID of the service to be enabled or disabled and a boolean value indicating whether to enable or disable the particular service. A True value indicates that the service should be enabled; while, a False value indicates that it should be disabled. The EML environment replies back with an "enable-ack" command that contains the ID of the addressed service and a boolean value indicating whether or not the operation was successful.

*getClients: service-ID*
*getClients-ack: service-ID, numberOfClients, True|False*

The "getClients" command is used to retrieve the number of clients connected to a particular service. The system's administrator initiates a "getClients" command sending as parameter the ID of the addressed service. The EML environment replies back with a "getClients-ack" command that contains the ID of the addressed service, the number of clients connected to the service, and a boolean value indicating whether or not the operation was successful. In case a false acknowledgment was returned, *numberOfClients* parameter would be equal to -1; whereas, in case a true acknowledgment was returned, *numberOfClients* would contain a positive integer value.

*grant: service-ID,*
*permission1:allow|deny;permission2:allow|deny;permissionN:allow|deny*
*grant-ack: service-ID,*
*permission1:True|False;permission2:True|False;permissionN:True|False*

The "grant" command is used to grant or revoke access rights or permissions to a particular service in the ecosystem. These permissions include but not limited to writing to disk, spawning a process, and accessing the network. The system's administrator initiates a "grant" command sending as parameters the ID of the addressed service and the list of permissions to be granted or revoked. Each permission name is followed by either *allow* to grant the permission or *deny* to revoke it. The EML environment replies back with a "grant-ack" command that contains the ID of the addressed service and the list of permissions together with a list of Boolean values indicating whether every particular permission was successfully granted or revoked.

*createReplica: service-ID, replicaServerIP*
*createReplica-ack: service-ID, replica-service-ID, True|False*

The "createReplica" command is used to create a replica for an existing service. The system's administrator initiates a "createReplica" command sending as parameters the ID of the service to be replicated and the IP of machine that is going to host the service replica. The EML environment replies back with a "createReplica-ack" command that contains the ID of the addressed service, the ID (auto-generated by the EML environment) of the service replica, and a boolean value indicating whether or not the service was replicated successfully. In case a false acknowledgment was returned, *replica-service-ID* parameter would be equal to -1; whereas, in case a true acknowledgment was returned, *replica-service-ID* would contain a positive integer value. In order to delete an existing replica, the "unbind" command can be used to disintegrate (disconnect) the service replica from the ecosystem.

*getInfo: service-ID*
*getInfo-ack: service-ID, XML-report, True|False*

The "getInfo" command is used to retrieve details about an existing service. The system's administrator initiates a "getInfo" command sending as parameter the ID of the service whose details are to be fetched. The EML environment replies back with a "getInfo-ack" command that contains the ID of the addressed service, a report in XML format, and a boolean value indicating whether or not the operation was successful. In case a false acknowledgment was returned, *XML-report* parameter would be equal to null; whereas, in case a true acknowledgment was returned, *XML-report* would contain an XML formatted message [7]. The specifications of the XML report are given below:

```
<report>
    <serviceID>23</serviceID>
    <serviceIP>192.168.1.20</serviceIP>
    <serviceWSDL>WSDL Description</serviceWSDL>
    <isEnabled>True|False</isEnabled>
    <isReplica>True|False</isReplica>
    <grantedPermissions>
        <permission>x</permission>
        <permission>y</permission>
        <permission>z</permission>
        …
        …
        …
    </grantedPermissions>
    <stamp>12/7/2011 08:15:21PM</stamp>
    <version>1.0</version>
</report>
```

In order to validate whether the XML report conforms to the specifications of the EML language, a DTD validator [8] is employed to verify the correctness of the grammar and syntax of the report. The DTD definition is given below:

```
<!ELEMENT report (serviceID, serviceIP, serviceWSDL,
isEnabled, isReplica, grantedPermissions, stamp, version)>
<!ELEMENT serviceID (#PCDATA)>
<!ELEMENT serviceIP (#PCDATA)>
<!ELEMENT serviceWSDL (#PCDATA)>
<!ELEMENT isEnabled (True|false)>
<!ELEMENT isReplica (True|false)>
<!ELEMENT grantedPermissions(permission*)>
<!ELEMENT permission (#PCDATA)>
<!ELEMENT stamp (#PCDATA)>
<!ELEMENT version (#PCDATA)>
```

As the EMU features a set of SAP procedures, they can be executed by the following EML command:

*executeSAP: service-ID, SAP-Procedure*
*executeSAP-ack: service-ID, True|False*

The "executeSAP" command is used to execute a Self-Adaptation Procedure (SAP). In effect, SAPs are built-in functions or procedures that are provided by the EMU to deliver self-adaptation properties to the digital ecosystem. The system's administrator initiates an "executeSAP" sending as parameters the ID of the service over which the SAP is to executed and the actual SAP procedure name. The

EML environment replies back with a "executeSAP-ack" command that contains the ID of the addressed service and a boolean value indicating whether or not the SAP was executed successfully.

## EML SAP – SELF-ADAPTATION PROCEDURES

Characteristically, a digital ecosystem should exhibit self-adaptation properties i.e. the ability to self-adapt and self-optimize according to the state of the ecosystem resources and to its execution environment [9]. The EML language provides a set of built-in procedures for delivering self-adaptation functions for the ecosystem allowing it to change its state based on the state of its operating setting such as increasing memory allocation, increasing disk quota, assigning more CPU cores and cycles, and reducing power consumption. SAPs are incorporated inside the Ecosystem Management Unit (EMU) and are executed through the Ecosystem Management Language (EML). For this purpose, a special engine is used to decode SAPs and transfer them accordingly to the operating system drivers layer which interfaces the computer hardware with the operating system. On the low-level, SAPs are instructions for the computer hardware to alter its configuration and behavior. Figure 3 depicts the SAP engine and its function.

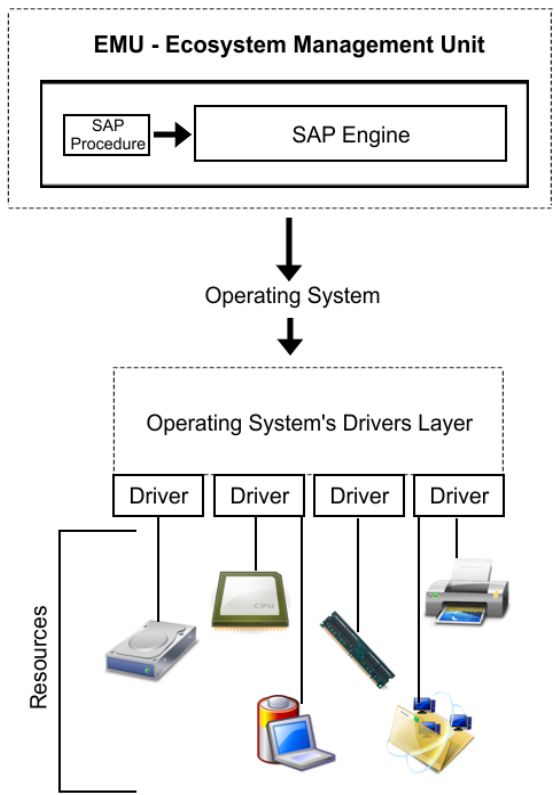

Figure 3. SAP engine

Practically, SAP procedures have a wide scope of applications; they can be brought down into several categories that are listed below:

Dynamic Power Management: SAPs in this category manage the electrical power that is supplied to the ecosystem, and are intended to reduce power consumption of certain nodes when they are in idle mode or not being used by any service or client.

**Dynamic Power Supply:** SAPs in this category represent the capabilities and management capacity of an uninterruptible power supply (UPS). The properties of the UPS device indicate when incoming power is trimmed or boosted, and the aggregated information of the power supply that comprises the ecosystem. Such type of SAPs can be used to switch to another source of power when a certain one fails, preventing ecosystem stoppages and maintaining the availability and uptime of the system.

**Dynamic CPU Overclocking:** SAPs in this category are directly connected to the computer system's basic input/output services (BIOS). Such type of SAPs can be used to boost-up the speed of certain machines that are hosting computationally-intensive services. As a result, dynamic performance can be achieved depending on the service being executed.

**Dynamic CPU Cores Allocation:** SAPs in this category manage CPU internal cores and are intended to dynamically allocate extra processor cores for multithreaded services that are receiving too much traffic. Additionally, in a symmetric multiprocessing operating system, processor affinity can be modified so that each task is allocated to a certain processor in preference to others.

**Dynamic Fans Allocation:** SAPs in this category manage fan devices in a server computer through the BIOS firmware. Such type of SAPs can be used to cool down an overloaded CPU or a machine that has been operating for a long time.

**Dynamic Disk-Space Allocation:** SAPs in this category manage computer physical secondary storage along with its associated mapped addresses. Such type of SAPs can be used to dynamically allocate additional disk-space for certain services requiring extra storage.

**Dynamic Memory-Space Allocation:** SAPs in this category manage computer system's primary memory along with its associated mapped addresses. Such type of SAPs can be used to allocate additional memory-space for certain services requiring more RAM space.

**Dynamic Network Bandwidth Allocation:** SAPs in this category control the behavior of a network adapter including the in and out traffic throughput, network bandwidth, and connection speed. Such type of SAPs can be used to allocate additional network bandwidth for certain services requiring higher network and Internet speed.

**Dynamic Printers Allocation:** SAPs in this category manage printer devices that are connected to the ecosystem network. They additionally define the configuration for a printer device including printing resolution, color, fonts, and orientation. Such type of SAPs can be used to switch between printers when one's ink is out or a paper is jammed or even when one is overloaded (busy) with printing jobs.

## EXPERIMENTS & RESULTS

In the experiments, an E-learning digital ecosystem model was built and tested. It comprises three layers: The presentation layer delivering the system's input and output interfaces; the service layer hosting all the system's services; and the data layer housing the system's data storage. The service layer is majorly composed of several web services ready to be consumed by users and client applications. They are but not limited to the "QUIZ" web service which issues exams and workouts; the "TUTORIAL" web service which represents a virtual interactive tutor; the "ENCYCLOPEDIA" web service which offers articles and extracts; and the "SEARCH" web service which helps students find documents, articles, handouts, and other learning materials. Figure 4 shows the diagram of the E-learning model under test.

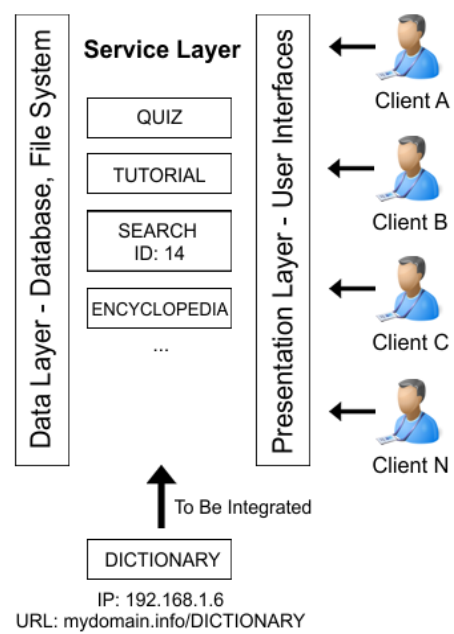

Figure 4. E-Learning model

For testing purposes, a new web service called "DICTIONARY" originally hosted on a server whose IP is equal to *192.168.1.6* and located on *mydomain.info/DICTIONARY* was integrated into the existing ecosystem using the *bind* command of the EML language. The service was integrated successfully and was assigned a random ID equals to 2. Then it was enabled using the *enable* command, and granted a disk access right using the *grant* command. Finally, information about this new service was retrieved using the *getInfo* command. Table I delineates the EML commands and their returned acknowledgments used in the test.

Table I.   EML Commands

| Operation | Issued by | EML Commands |
|---|---|---|
| Integrate service | *Admin* | *bind: mydomain.info/DICTIONARY, WSDL* |
| | *DICTIONARY Service* | *bind-ack: 2, True* |
| Enable service | *Admin* | *enable: 2, True* |
| | *DICTIONARY Service* | *enable-ack: 2, True* |
| Grant disk access | *Admin* | *grant: 2, disk:allow* |
| | *DICTIONARY Service* | *grant-ack: 2, disk:True* |
| Get Information | *Admin* | *getInfo: 2* |
| | *DICTIONARY Service* | *getInfo-ack: 2, XML-report, True*<br><br>*XML-report:*<br><report><br>  <serviceID>2</serviceID><br>  <serviceIP>192.168.1.6</serviceIP><br>   <serviceWSDL>WSDL-Description</serviceWSDL><br>  <isEnabled>True</isEnabled><br>  <isReplica>False</isReplica><br>  <grantedPermissions><br>    <permission>Disk Access</permission><br>   </grantedPermissions><br>  <stamp>12/7/2011 08:15:21PM</stamp><br>  <version>1.0</version><br> </report> |

Furthermore, self-adaptation was tested. In fact, the "SEARCH" web service has an ID equals to 14 and is of high-demand as it is subject to many client requests. Often, an increase in the number of users can lead to an increase in traffic and thus can impose a bandwidth problem on the communication lines. For this reason, a SAP procedure was used to allocate more network bandwidth and Internet resources to the "SEARCH" web service. Below is the EML command with the SAP procedure.

*executeSAP: 14, IncreaseNetBandwidth()*
*executeSAP-ack: 14, True*

## CONCLUSIONS & FUTURE WORK

This paper presented a specification for a management language called EML for automating the control and management of service components connected to a digital ecosystem. It relies on proprietary syntax to format its instructions mainly composed of managerial commands. It additionally supports Self-Adaptation Procedures called SAP that allow the ecosystem to self-adapt according to its computing environment. EML commands are interpreted and processed by an internal unit called EMU short for Ecosystem Management Unit. All together, they automate the manageability of the ecosystem resources by allowing system's administrators to control, monitor, and report on the ecosystem service components.

As future work, the EML is to be extended so that it supports additional administrative commands and a larger library of SAP procedures allowing further control over the various entities of the digital ecosystem.

## ACKNOWLEDGMENT

This research was funded by the Lebanese Association for Computational Sciences (LACSC), Beirut, Lebanon under the "Digital Ecosystem Research Project – DERP2011".

## REFERENCES

[1] Dini, P., "Towards Business Cases and User-Oriented Services in Digital Business Ecosystems", Workshop on Needs and Requirements of Regions, Brussels, 2005.

[2] Nachira, Nicolai, Dini, Le Louarn, & Leon, "Digital Business Ecosystems", European Commission, 2007.

[3] Ferronato, P., "Architecture for Digital Ecosystems, beyond Service Oriented Architecture", Digital Ecosystems and Technologies Conference, DEST '07, 2007.

[4] Corallo, A., Passiante, G., & Prencipe, A., The Digital Business Ecosystem, Edward Elgar Pub, 2007.

[5] Hoque, F., "e-Enterprise: Business Models, Architecture, and Components", Cambridge University Press, 2000.

[6] Official OASIS Standard, OASIS Reference Model for Service Oriented Architecture 1.0, http://docs.oasis-open.org/soa-rm/v1.0/soa-rm.pdf, 2006.

[7] W3C, Extensible Markup Language XML Specifications, http://www.w3.org/XML/, 2003.

[8] Bex, G., Neven, F., & Bussche, J., "DTDs versus XML schema: a practical study", Proceedings of the 7th International Workshop on the Web and Databases collocated with ACM SIGMOD/PODS., 2004.

[9] Hadzic, M., Chang, E., & Dillon, T., "Methodology framework for the design of digital ecosystems, Systems, Man and Cybernetics", ISIC IEEE International Conference, pp7–12, 2007.